\documentclass[useAMS,usenatbib]{mn2e}
\usepackage{epsfig}
\newcommand{\be}{\begin{equation}}
\newcommand{\ee}{\end{equation}}

\newcommand {\bea}{\begin{eqnarray}}
\newcommand {\eea}{\end{eqnarray}}
\newcommand {\lI}{\lambda_1}
\newcommand {\lII}{\lambda_2}
\newcommand {\lIII}{\lambda_3}

\title[Why does the clustering of haloes depend on their formation history?]{Why does the clustering of haloes depend on their formation history?}
\author[H.B. Sandvik, O. M\" oller, J. Lee and S.D.M. White]{H.B. Sandvik$^{1}$\thanks{E-mail:
    sandvik@mpa-garching.mpg.de (HBS); ole@mpa-garching.mpg.de (OM)}, O. M\" oller$^{1}$\footnotemark[1], J. Lee$^{2}$ and
  S.D.M. White$^{1}$ \\
$^{1}$ Max Planck Institut f\" ur Astrophysik, D-85741 Garching, Germany \\
$^{2}$ School of Physics and Astronomy, FPRD, Seoul National
  University, Seoul 151-742, Korea}
\begin{document}



\maketitle

\label{firstpage}

\begin{abstract}
  We discuss in the framework of the excursion set formalism a recent
  discovery from N-body simulations that the clustering of haloes of
  given mass depends on their formation history. We review why the
  standard implementation of this formalism is unable to explain such
  dependencies, and we show that this can, in principle, be rectified
  by implementing in full an ellipsoidal collapse model where collapse
  depends not only on the overdensity but also on the shape of the
  initial density field. We also present an alternative remedy for
  this deficiency, namely the inclusion of collapse barriers for
  pancakes and filaments, together with the assumption that formation
  history depends on when these barriers are crossed.  We implement
  both these extensions in a generalised excursion set method, and run
  large Monte Carlo realisations to quantify the effects. Our results
  suggest that effects as large as those found in simulations can only
  arise in the excursion set formalism if the formation history of a
  halo does indeed depend on the size of its progenitor filaments and
  pancakes.  We also present conditional distributions of progenitor
  pancakes and filaments for low-mass haloes identified at present
  epoch, and discuss a recent claim by Mo et.al. that most low-mass
  haloes were embedded in massive pancakes at $z\sim 2$.
\end{abstract}


\section{Introduction}\label{introduction}
The excursion set formalism \citep{PS74,bond91,Bower91,LC93,MW96} has
been remarkably successful in describing several characteristics of
the halo population in numerical simulations of structure formation,
such as their unconditional and conditional mass functions, their
merger rates, the halo bias and more. In general terms the approach
implies picking a ``particle'' at random and smoothing the linear
density field over ever smaller spheres around it, until the criterion
for collapse at some redshift $z$ is satisfied. The mass in the sphere
is then identified as that of the collapsed object to which the
particle belongs. This criterion is typically a constraint on the
linear density contrast obtained from a collapse model, either
spherical or ellipsoidal. The spherical collapse model gives a simple
critical value for collapse $\delta_c \approx 1.686$ above which a
patch is said to have collapsed at present time \citep{PS74}. For the
most popular implementations of the ellipsoidal model this is modified
only by making the collapse criterion scale-dependent
$\delta_c(\sigma(R))$, because shape parameters are approximated by
their expectation values \citep{SMT01, ST02}. Thus determining whether
a patch has collapsed or not requires knowledge only of the density
contrast $\delta$ and the smoothing scale $R$.

When smoothing the density field a specific choice of filter shape
must be used. An intuitively appealing choice is the top-hat filter
for which both visualisation of the problem and the assignment of a
mass to each collapsed region are straightforward. However, since the
$k-$modes of the linear perturbation field are independent, it is
mathematically more convenient to use a sharp $k-$space filter, since
varying resolution then corresponds to a Markov random walk in the
density contrast. As it turns out, this is also the filter which
best reproduces the shape of mass functions seen in N-body
simulations \citep{percival01}, although a different filter then needs
to be employed to assign a halo mass to each smoothing scale. It is
surprising that this simple formalism, based only on the
linear density field, is so successful in explaining many of the
features seen in N-body simulations.

Despite its successes the approach clearly has shortcomings. Perhaps
the most studied arises from the inapplicability in detail of the
physical model which underlies it. This stems from the fact that few
particles actually lie at peaks of the smoothed initial density
field. Most lie somewhere on the peripheral slopes of the peak
associated with the halo they will actually belong to at the target
redshift.  As a result, there is a relatively poor match between the
halo mass predicted for individual simulation particles by the random
walk associated with the linear density field and the mass of the
actual halos in which they find themselves as a result of the
nonlinear evolution \citep{bond91,White96,SMT01}. The excellent agreement
of the statistics of simulated halos with excursion set theory
predictions thus does not extend to good object-by-object agreement.
\citet{BM96} devised an algorithm for explicitly taking account of
the non-locality of the collapse process by assigning a surrounding
``patch'' to each ``peak'' in the initial conditions. This adds
considerable complexity, however, and as a result is rarely
used. Despite this problem, the excursion set approach has proved
remarkably useful.

Another weakness which has received recent attention
(\citealt{Wang06,ST04}, see also \citealt{White96}) is the fact that in its simplest and most used
approximation, the approach does not allow correlations between halo
properties defined by the character of the random walks on opposite
sides of the barrier crossing. One example where such correlations are
seen in simulations is the discovery by \citet{GSW05} (GSW05) that old
haloes are more clustered than young haloes of similar mass, or, more
generally, that the clustering of halos depends on their formation
history as well as on their mass \citep[][CGW06]{CGW06}. This is also
manifest as a dependence of clustering on properties such as halo
concentration or substructure fraction \citep{Wechsler06}.  All these
effects require some degree of correlation between environment density
and halo formation history. In the standard 1-dimensional excursion
set approach such correlations are impossible by construction.

There are two ways to extend the excursion set approach to address
these effects. For a multidimensional random walk of the kind
pioneered by \citet{CL01}(CL01) the extra variables can carry
information across the barrier. A particle with a high environment
density can cross at a different point in terms of these extra
variables than a low density particle, and this can then be reflected in
the random walk at smaller scales, hence in the formation history of
the particle's halo. A second way to account for such correlations
arises if the ``shape'' of a halo's formation history depends on
\emph{different} barriers than does its mass.  For example, if we
allow for the possibility that the assembly history of a halo could be
significantly impacted by the size of its progenitor pancakes and
filaments, an interesting scenario appears: it is then possible that
the progenitor pancake scale is \emph{larger} than the present halo
scale, meaning the pancake barrier crossing lies between the halo
collapse crossing and the scale defining the environment.  This makes
possible some correlation of formation history with environment
density for halos of given mass, as seen in simulations. In this paper
we will study both these possibilities in detail.

In order to allow for multi-scale correlations we implement the 6D
Markov approach pioneered by CL01 using realisations of the
deformation tensor rather than merely the density contrast, thereby
utilizing the full shape information.  We introduce a new set of shape
parameters which allow us to construct a halo collapse barrier which
reproduces very well the dynamical collapse model of
\citet{BM96}(BM96).

To be able to discuss the second possibility laid out above, we use
the same dynamical collapse model to construct barriers corresponding
to filaments and sheet formation. Such barriers were first presented
in a recent paper by \citet{shen06}, who used them, among other
things, to provide interesting analytic estimates of the mass
fractions in pancakes, filaments and haloes at any given
time. Although they used the average shape approximation, something we
will avoid here, their barriers could in principle also have been used
for parts of the discussion in this paper. An alternative approach
using the Zel'dovich approximation to discuss pancake formation was
recently suggested by \citet{Lee06}.

Our paper is laid out as follows. We first review the standard
implementation of the excursion set method with flat and moving
barriers and demonstrate why this is unable to reproduce the
correlations discussed above. We then explain the generalisation of
CL01 to a random walk in the deformation tensor. We show how choosing
a particular linear combination of the eigenvalues allows the
behaviour of the ellipsoidal collapse model to be successfully
captured by barrier functions for haloes, filaments and sheets. We
then run large samples of Monte Carlo realisations and analyse the
results. We will conclude that it is possible to understand the
clustering dependence on formation history if we make the plausible
assumption that the latter depends on the masses of progenitor
pancakes and filaments.

\section{The excursion set approach}\label{theory}
\subsection{A random walk with a flat barrier}\label{flatBarrier}
The spherical collapse model identifies a linear theory density
contrast, $\delta_c(z) = \delta_c /D(z)$,
extrapolated to present epoch, which corresponds to collapse at
redshift $z$. Here $\delta_c \approx 1.686$ \citep{PS74} and $D(z)$ is
the linear theory growth function normalised to $1$ at $z=0$. This
equals $1/(1+z)$ in the special case of an Einstein de Sitter
universe. 

The traditional excursion set approach
entails picking a random position (particle) and  
smoothing the linear density field, $\delta$,  over ever
smaller spheres around this point until this spherical collapse
criterion is fulfilled ($\delta \ge \delta_c(z)$). 

The density perturbation smoothed over a scale $R$ is given by 
\be
\delta_R(x) = \int \delta(x') W_R(x - x')d^3x', 
\ee 
where $W_R(x)$ is the scale-dependent smoothing filter.  The variance
of the density field on scale $R$ is given by 
\be 
\sigma^2(R) = \langle | \delta_R(x) |^2 \rangle = { 1 \over 4 \pi^2} 
\int p(k)\tilde{W}^2_R(k) k^2 dk, 
\ee 
where $\tilde{W}_R(k)$ is the fourier transform of the filter
function. We are in principle free to choose this filter function, and
the top-hat filter is frequently used for visualisation purposes,
although computationally it is awkward since the added modes are
correlated as we go to smaller scales.  If instead a hard $k-$space
filter is used (with an upper $k-$space cutoff $k_c = 1/R$), the modes
added by going to a smaller scale are statistically independent of the
modes already included. Thus the density contrast performs a Markov
random walk, and the entire problem can be described by such a random
walk with an absorbing constant barrier, where $\sigma^2$ is the
pseudo-time variable.

A particle trajectory will thus gradually carry out a random walk as
$\sigma^2$ increases until the barrier is crossed at some particular
value of this variance.  This value corresponds to a smoothing scale
$R$, and the mass of the corresponding halo is usually assigned using
the top-hat formula (independent of the filter actually used for the
smoothing);
\be
M = {4 \over 3} \pi \bar{\rho} R^3.
\ee

For the flat barrier, the problem of a Markov random walk with an absorbing
barrier is analytically tractable. The first crossing distribution
is given by \citep{chandrasekar43}  
\be
\nu f(\nu) =  \left({\nu\over 2 \pi}\right)^{1/2} \exp{\left(-{\nu\over 2} \right)}
\ee
where $\nu = \delta_c^2 / \sigma^2$. 

The mass functions $n(m,z)$ of dark haloes are obtained from the
density distribution function of up-crossings $f(\nu)$, through
\be
\nu f(\nu) \equiv m^2 {n(m,z) \over \rho} {d \ln m \over d \ln \nu}.
\ee

It is important to realise that only the first crossing of a barrier
is of any importance. Any subsequent crossings of the \emph{same}
barrier are irrelevant since collapse has already occurred at this
redshift on a larger scale. However, an earlier redshift corresponds
to a higher barrier, and the first up-crossing of this higher barrier
corresponds to the largest scale on which collapse has occurred at
the earlier redshift. 

The conditional mass functions, defined as the distributions of
progenitor mass at an earlier redshift for given final halo mass, are
related to this two barrier problem. In the flat barrier case they are
given by a simple extension of the unconditional mass function.

Although the excursion set approach cannot tell us about the
substructure of a halo at any particular redshift, it does tell us
something about how mass was \emph{added} to the halo as it grew.
This is possible since we can follow the mass of the halo's main
progenitor with redshift, and thereby study its merger history
\citep{LC93}.  This allows us to identify the formation redshift $z_f$
as the redshift at which we can first identify a progenitor with more
than half the final mass.

\subsection{The moving barrier}\label{movingBarrier}
\citet{SMT01}(SMT01) were the first to use the ellipsoidal collapse
model of BM96 in the excursion set approach. In this model collapse
depends not just on the density field, but on eigenvalues of the tidal
shear tensor.  However, rather than drawing realisations of the shear
tensor, and thereby including full shape information, SMT01 used
expectation values of the shape parameters as a function of
$\sigma$. They were then able to construct a collapse barrier whose
value depended on the variance, $\sigma^2$, thus on mass. Since variance is
the time-like variable, this has been labelled a ``moving'' barrier.

With this shortcut, the approach, with a sharp $k$-space filter,
was highly successful in reproducing the shapes of the halo mass
functions seen in N-body simulations. SMT01 provided a convenient analytic
approximation to the first-crossing distribution corresponding to
their moving barrier, but the symmetry which allows for compact
representation of the conditional mass functions in the flat barrier
case is lost. 

As an extension to this work \citet{shen06} recently provided moving
barriers which can represent collapse along one, two and three axes,
corresponding to the formation of sheets, filaments and haloes, an
effort which we replicate and extend later in this paper.

\subsection{Correlations between clustering and formation}\label{correlations}
\begin{figure}
\centerline{\epsfxsize=8.0cm\epsffile{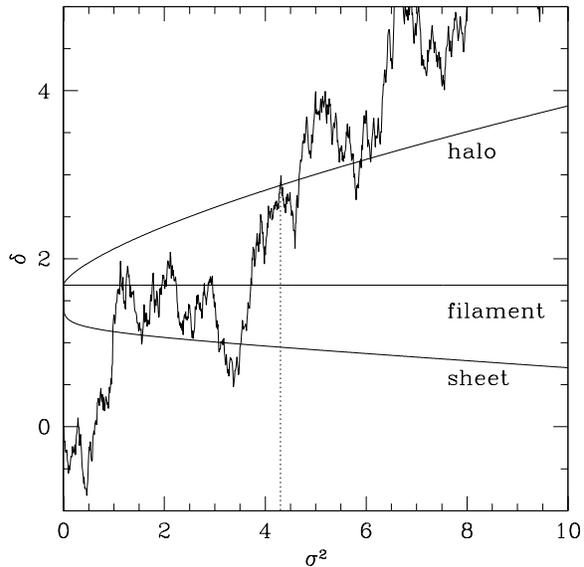}}
\caption{Example of a 1-dimensional random walk (jagged line) together
with the average barriers for sheets, filaments and haloes at
$z=0$. The variance at which first up-crossing occurs determines the
mass of each structure element in which this specific particle is
embedded. It is clear that each halo identified at $z=0$ is contained
in a larger mass filament which is in turn contained in an even larger
pancake. The environment density of the halo depends on values of
$\delta$ on yet larger scales, nearer to $\sigma = 0$, whereas, in the
usual formulation, its
formation history is determined by the walk's structure on small
scales to the right of halo crossing.}
\label{randomWalk}
\end{figure} 

It is easy to see why the usual excursion set approach precludes
correlations between large-scale environment and formation history for
haloes of \emph{given} mass \citep{White96}. Take the walk in figure \ref{randomWalk}
as an example. The walk crosses the $z=0$ halo barrier at
$\sigma_h^2=4.15$, and is so is identified as belonging to a halo of
mass $M(\sigma_h)$. The density of the large-scale environment for
this particle can be identified using the values of $\delta$ near the
origin, $\sigma^2 \ll \sigma_h^2$ (and $\delta \ll \delta_c$), whereas
its formation history (at least, according to the usual definitions)
depends on the behaviour of the random walk to the right of the
barrier crossing at $\sigma_h^2=4.15$.  Since, by assumption, all
haloes of given mass cross the $z=0$ halo collapse barrier at the same
variance, the Markov nature of the random walk prevents any
correlation between the two sides, between environment and formation
history. In other words, old haloes should cluster in exactly the same
fashion as young ones.

In the following we discuss two ways in which the excursion set
approach can be extended to accommodate correlations between clustering
and formation history. In the first, a multi-dimensional
implementation of the random walk is combined with a non-spherical
collapse barrier, as pioneered by \citet{CL01}. This implies that halo
collapse at given mass is associated with a single constraint on a set
of variables, in their case $\{\delta, r\}$.  Formulated differently,
the effective barrier for $\delta$ depends on the value of the shape
parameter $r$.  At barrier crossing $\delta$ and $r$ can thus have
different values for different halos (though they are perfectly
correlated) and these different values can, in principle, correspond
to different distributions of \emph{both} formation history and 
large-scale environment. A correlation between environment and
formation history thus becomes possible.

\begin{figure}
\centerline{\epsfxsize=8.0cm\epsffile{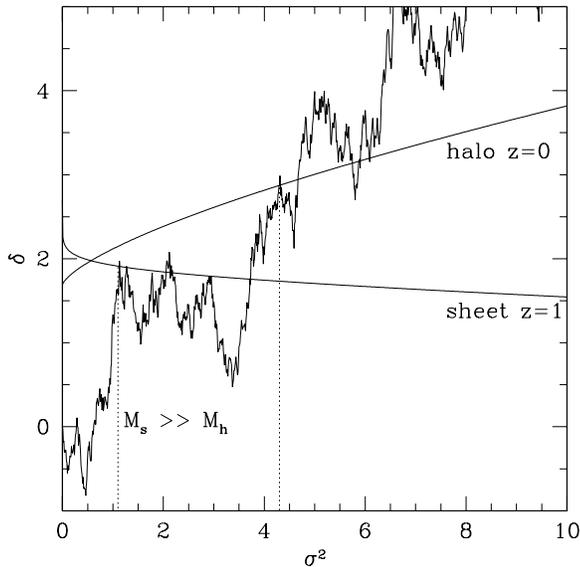}}
\caption{The random walk in figure (\ref{randomWalk}) with the $z=0$
halo barrier and $z = 1$ sheet barrier. This trajectory corresponds to
a halo whose $z=1$ progenitor was part of a pancake with mass
exceeding the $z=0$ halo mass. Assuming the assembly history of the
final halo to be influenced by the mass of this early pancake provides
a possible explanation for the dependence of clustering on formation
history, since both the large-scale density and the mass of the early
pancake correspond to points on the random walk to the left of the
barrier crossing which defines the halo's $z=0$ mass.}
\label{AHhalo}
\end{figure} 

A second possibility is to stipulate that formation history of a halo
somehow depends not only on its progenitor haloes, but on progenitor
pancakes and filaments as well. These structures are characterized by
collapse along one or two axes, respectively, and can be represented
in the excursion set approach by different collapse barriers, as first
suggested by \citet{shen06}. Figure \ref{randomWalk} compares the
average shapes of such barriers at $z=0$. The general trend is clearly
very different as we go to smaller scales; the halo barrier increases
with $\sigma^2$, the filament barrier is constant, and the pancake
barrier actually decreases as we go to smaller scales. Since going to
higher redshift simply scales the barriers up and to the right, it is
possible, even probable for small masses, that the mass of a progenitor
pancake at some early time is larger than the final mass of the halo.
Figure \ref{AHhalo} shows an example of such an occurrence, where the
progenitor of a $z=0$ halo was part of a larger mass pancake at $z =
1$.  The idea that halo formation history could somehow depend on a
different barrier than halo mass, is crucial, since there is every
reason to expect a clear correlation between properties defined by the
behaviour of the random walk on the {\emph same} side of the barrier
crossing which defines halo mass.

We have identified two independent extensions which can, in principle,
accommodate the kind of correlation between environment and formation
history that we seek. To examine whether these extensions can provide
a quantitative explanation of the numerical results, we need to
implement them in a generalised excursion set approach.  This is the
subject of the next section.


\section{The generalized excursion set method}
\subsection{The 6D Random Walk}
The general non-spherical excursion set approach with a hard $k$-space
filter involves a random walk in the deformation tensor (the 2nd
derivative tensor of the peculiar gravitational potential), 
\be
d_{ij} = \Phi_{,ij},
\ee
rather than simply in the density contrast alone (which is the trace
of this tensor). CL01 devised an algorithm for this:

The deformation tensor $d$ can be simulated by drawing six independent
Gaussian variables $\{y_1..y_6\}$ with dispersion $\sigma_0$, and
using the following linear transformation.
\bea
d_{11} &=& -{1\over 3}\left(y_1 + {3\over \sqrt{15}}y_2 + {1\over
  \sqrt{5}}y_3\right) \\
d_{22} &=& -{1\over 3}\left(y_1 - {2\over \sqrt{5}y_3}\right)\\
d_{33} &=& -{1\over 3}\left(y_1 - {3\over \sqrt{15}}y_2 + {1\over
  \sqrt{5}}y_3\right) \\
d_{12} &= &d_{21} = {1 \over \sqrt{15} y_4} \\
 d_{23} & = &d_{32} = {1 \over  \sqrt{15}} y_{5} \\
 d_{13} & = & d_{31} = {1\over \sqrt{15}} y_6.
\eea
This transformation satisfies the correlations of the deformation
tensor as shown in \citet{BBKS86}.

The random walk proceeds by drawing new values from the Gaussian
distribution, and adding to ${y_i}$, until we have a walk of $N$
steps. To each step $n$ can be assigned a dispersion $\sigma^2 =n
\sigma_0^2$. 

For each step of the walk the deformation tensor can be diagonalised
to find the three eigenvalues $\{\lambda_1, \lambda_2,\lambda_3\}$. 
This set of eigenvalues, along with a non-spherical collapse model,
determine whether collapse has occurred by a given
redshift. Although the eigenvalues could be drawn directly from the probability
distribution function of \citet{doroshkevich70},
\bea
p(\lambda_1, \lambda_2, \lambda_3) &=& {15^3 \over 2^3}{1 \over
  \sqrt{5}\pi\sigma_M^6} \exp\left( - {3 I_1^2 \over \sigma_M^2} + {15
  I_2 \over 2 \sigma_M^2} \right) \\
& & \times (\lambda_1 - \lambda_2) (\lambda_2-\lambda_3) (\lambda_1 - \lambda_3),
\label{doroshkevich}
\eea
the above approach is significantly faster.  

CL01 chose to avoid the diagonalisation process altogether by
expressing the collapse criterion in terms of $\delta$ and of a
variable $r^2$ directly calculable from the $y_i$s through two
rotational invariants. This speeded up their calculations and allowed
a nice visualisation of the problem.  We discovered that this
simplification results in a suboptimal representation of the fully
3-dimensional ellipsoidal collapse model, and we choose instead to use
diagonalisation of the deformation tensor to obtain the the
eigenvalues ${\lambda_1,\lambda_2 , \lambda_3 }$. The change of
variables ${\lI,\lII,\lIII} \rightarrow {\delta,v,w}$ given by
\bea
\delta &=& \lI + \lII + \lIII \\
v & = & -\lI + \lII\\
w & = & -\lI -\lII +2 \lIII
\eea
allows simple and accurate modelling of the behaviour of the
ellipsoidal collapse model. This choice is discussed in detail in
Appendix \ref{invariants}.   

\subsection{Collapse Barriers}\label{barriers}
The spherical collapse barrier is found by applying the spherical
collapse model and identifying the value of the initial density
contrast, \emph{linearly extrapolated} to present time, which
corresponds to collapse at redshift $z$. It thus represents a constant
(flat) barrier for the random walk. In the ellipsoidal collapse
model, collapse is determined by the shape of the initial patch as well
as its overdensity and this results in a barrier which depends on all
3 eigenvalues $\lambda_i$ rather than just on their trace $\delta$. 

To fully capture this dependency we use the following barrier shape to
represent halo collapse. 
\be
S_{H}(v,w,z) = {\delta_c \over D(z)} \left[1 + \alpha_1 \left({v
  D(z)}\right)^{\alpha_2}+\alpha_3 \left({w D(z)}\right)^{\alpha_4}
\right]
\ee
where $\delta_c = 1.686$ is the critical value for spherical collapse
at the present epoch, $D(z)$ is the linear theory growth factor
normalised to $1$ at $z=0$, and
\be
\alpha_1\approx 0.2809, \alpha_2 \approx 1.3557, \alpha_3 \approx
0.070, \alpha_4 \approx 1.41205
\ee

The coefficients here were found by fitting to the collapse values of
$8000$ walks (stepsize $\Delta \sigma^2 = 0.025$) for which we used the full numerical
solution to the ellipsoidal collapse model of BM96 (see
appendix). This interpolation formula provides a near \emph{perfect}
fit to the true barrier obtained using the numerical solution, and it
allows for a thorough investigation of the limitations and accuracy of
the excursion set approach.  We want to stress that our barrier
was {\emph not} found by fitting to N-body mass functions, but rather by
requiring an accurate representation of the ellipsoidal collapse model
of \citet{BM96}. 

We also present barrier functions representing collapse along
one and two axes.  For sheets we have
\be
S_S(v,w,z) = {\delta_c \over D(z) }\left[1 - \alpha_1 (v D(z))^{\alpha_2} -
\alpha_3 (w D(z))^{\alpha_4} \right]
\ee
where $\alpha_1 = 0.3748, \alpha_2 = 0.2399, \alpha_3 = 0.003237,
\alpha_4 = 2.4187$. The formula for the filaments is slightly more
complicated since it needs to reproduce halo-like behaviour for two
large axes, and sheet-like behaviour for two short axes.
\be
S_F(v,w,z) = {\delta_c \over D(z)}\left[1 + \alpha_1 y D(z) \left(1 +
  \alpha_2 (x D(z))^{\alpha_3} \right)\right] 
\ee
where $\alpha_1 = 0.1104, \alpha_2 = 0.01641, \alpha_3 = 1.463$, and
$x = w+3v, y = w - 3v$. This change of variables is meaningful, since
the condition of perfect triaxiality, for which the behaviour of the
second axis is nearly identical to the spherical case \citep{shen06},
translates to $w = 3 v$.  Using the expectation values found in
Appendix \ref{invariants}, we can translate these barriers into
``moving barriers'' similar to \citet{shen06}. These average barriers
are the ones plotted in figure \ref{randomWalk}.

\section{Monte Carlo Simulations}
To address the questions posed in section \ref{theory} we simulate a
large ensemble ($\sim 10^6$) of random walks. For each simulated
trajectory we output the density contrast on large scales along with
the first up-crossings (actually the corresponding masses) of the
halo, filament and sheet barriers (sec. \ref{barriers}) for a range of
different redshifts. We also store the nominal formation redshift for
the $z=0$ halo, as well as the values of the auxiliary variables $v$
and $w$ at collapse.  In the following we analyse the results and we
try to understand their implications.

\subsection{First crossing distributions}
\begin{figure}
\centerline{\epsfxsize=8.0cm\epsffile{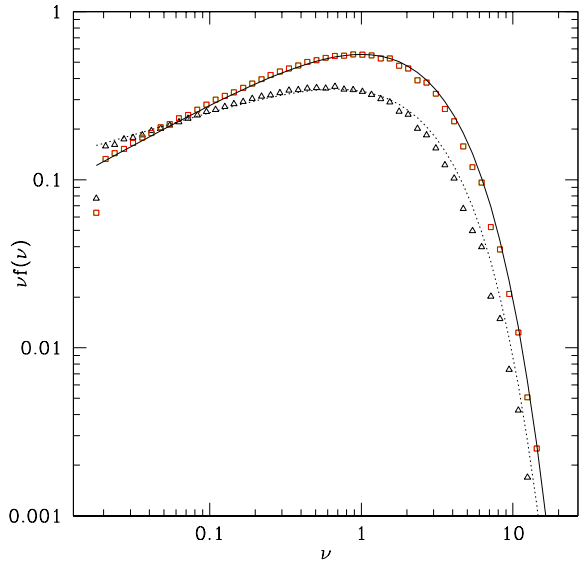}}
\caption{First crossing distributions obtained using the spherical
  collapse barrier (red squares) and the ellipsoidal collapse barrier
  presented in this paper (black triangles). The solid line is the
  analytical Press-Schechter solution, and the dotted line is the
  Sheth-Mo-Tormen approximation.}
\label{massFunctions}
\end{figure}

Figure (\ref{massFunctions}) shows the simulated first crossing
distributions for the halo barrier introduced in section
\ref{barriers} (squares) and for the spherical collapse barrier
(triangles). These are compared with the corresponding analytic
predictions from \citet{PS74} and SMT01. It should come as no
surprise that the distribution for our ellipsoidal collapse model is
well fit by the SMT01 analytic formula, since this was obtained by
fitting to a ``moving barrier'' version of our approach. Note that
their final mass function contained an additional scaling
parameter $a$, which allowed for better agreement with simulations. 
This degree of freedom is absent in our approach, though one might
argue that it should, perhaps, be reintroduced when comparing our
predictions to numerical results obtained with any particular (and to
some extent arbitrary) definition of the boundary and thus the mass of a
halo.

\subsection{Correlations between large and small scales}

As laid out in section \ref{correlations} a multi-dimensional random
walk with a shape-dependent barrier allows, in principle, for
correlations between the environment density on large scale and the 
point at which the collapse barrier is crossed.

In the following we use our Monte Carlo sample of random walks to
quantify this connection between large-scale overdensity, point of
barrier crossing and, potentially, halo formation redshift.

\subsubsection{Environment density and point of barrier 
crossing}\label{barrierCrossing}

From our ensemble of walks we pick out those in the upper and lower
$10\%$ tails of the environment overdensity distribution (estimated as
$\delta(\sigma^2 = 0.5)$).
This represents, in a statistical sense, our most and least
clustered haloes. We then compare the distributions of $v/\sigma$ and
$w/\sigma$ at barrier crossing for the two sets. Concentrating on ratios removes the 
obvious dependence on $\sigma$, providing shape parameters which allow
us to compare positions of barrier crossing in a way which does not
depend on overall barrier scale.

For the ellipsoidal collapse barrier we find a clear albeit small
correlation between environment and point of crossing. Both $w$ and
$v$ are on average higher for walks with high environment density.
The average value of $v/\sigma$ is roughly $7-8\%$ larger for the
walks with high environment density than for walks in the low density
tail. The dependence on $w$ is somewhat less pronounced ($5-6\%$
difference between the two tails). As expected no such
effects are found when we use the spherical collapse boundary.

It is worth noting that this effect is stronger for low values of
$\sigma^2$ at barrier crossing (i.e. for massive haloes) than for less
massive objects. This can be understood by recalling that the random
walk is a diffusion process, so that correlations with values at
``early times'' (i.e. large scales, small variance values) are
gradually washed out as ``time/distance'' increases (i.e. as variance
increases). Throughout this work we define the environment by the
overdensity at a fixed scale $\sigma^2 = 0.5$, roughly corresponding
to $R \sim 10 h^{-1}$Mpc in the concordance $\Lambda$CDM cosmology we are
adopting. With this definition it is not unexpected that environment
correlates less with properties at barrier crossing as we go to
smaller scales (larger $\sigma^2$ and smaller masses). This gives a
first hint that the 6D excursion set approach will be unable to
explain the effects measured by \citet{GSW05}, since these are largest
for low mass haloes, and are virtually non-existent for the most
massive haloes.

\subsubsection{Clustering and Patch Shapes}\label{patchShapes}

It is perhaps more interesting to study environmental effects on
collapse properties we are familiar with. For the purpose of this
analysis we estimate the ellipticity and prolateness of the initial
density field associated with a halo by
\be 
e =(-\lambda_1 + \lambda_3)/\sigma  
\label{ellipticity}
\ee 
and 
\be
p = (\lambda_1 - 2\lambda_2 + \lambda_3)/\sigma 
\label{prolateness}
\ee 

\begin{figure}
\centerline{\epsfxsize=9.0cm\epsffile{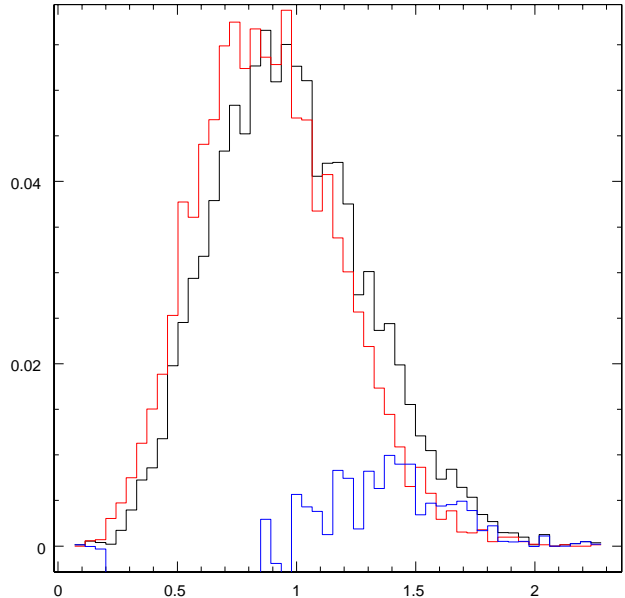}}
\caption{ Histogram of $e$ values at first crossing for the 10\%
highest (black) and 10\% lowest (red) tails of environment density
(estimated at $\sigma^2=0.5$) for the ellipsoidal collapse barrier. This
plot refers to haloes more massive than the characteristic mass
$M_*$. Although the effect is slight, the difference between the
distributions is statistically highly significant. The values of $e$
are on average $\sim 10\%$ larger at the point of first crossing for
walks in the high environment density tail than for walks in the
corresponding low density tail.}
\label{e_ellipsoidal}
\end{figure} 

We find that there is a weak correlation between environment density
and the pre-collapse ellipticity of haloes (see eqn.(\ref{ellipticity}
for the relevant definitions).  Again we see that the effect fades as
we go to smaller masses. Figure (\ref{e_ellipsoidal}) shows histograms
of $e$ values for the $10\%$ tails of the distribution in environment
density for haloes with $M > M_*$. A corresponding plot for haloes
with $M < 10^{-2}M_*$ shows no discernible difference between the two
distributions.  Roughly speaking, for massive haloes, those with the
10\% highest (lowest) environment densities have ellipticities which
are systematically 5\% higher (lower) than those of typical
haloes. Systematic effects with prolateness are also clear, albeit
smaller, indicating that haloes in dense environments have a wider
than typical range of prolateness values - in agreement with the
results for the ellipticities.

These results show shows that haloes in dense regions originate from
more elliptical initial patches than less clustered objects of
similar mass. It is tempting to extrapolate this conclusion to apply
to the actual ellipticity of the collapsed, quasi-equilibrium
halo. This is speculative, however, since we do not know how (or even
if) the final shape of a halo is related to the shape of its initial
patch.

\subsubsection{Environment Density and Formation Redshift}\label{formationRedshift}

In sections (\ref{barrierCrossing} and \ref{patchShapes}) we
established that there is a small but clear correlation between
environment overdensity and the point of barrier crossing, thus with
the shape of the collapsing patch.  The point of barrier crossing
clearly influences the properties of the random walk at larger
variance, hence on smaller mass scales, so we may expect a correlation
between environment density and formation redshift.  This is precisely
what was detected in N-body simulations by \citet{GSW05}. They show
that low-mass haloes with high formation redshifts are significantly
more clustered than similar mass but late-forming objects.
Hence they sit in denser environments.

We want to compare our results as directly as possible with the
results of GSW05. 
They look at a sample of haloes with particle numbers in the
range 100-200. Each particle has a mass of $8.6 \times 10^8 h^{-1}
M_\odot$, so this corresponds roughly to $\sim 1-2 \times 10^{11} h^{-1}
M_\odot$. In terms of the characteristic mass $M_* = 6.15\times
10^{12} M_{\odot}$ this is roughly $M = 0.14-0.28 \times M_*$.

We try to replicate this by taking similar samples from our random
walks. We use an effective power spectrum index of $n=-2$, and look at
two mass bins, with masses in the ranges $M = [M_*,2M_*]$, and $M =
[0.02M_*,0.1M_*]$. To improve statistics we use larger bins than GSW05
and we compare environment overdensity for haloes in the 20\% (rather
than 10\%) tails of the distribution of formation redshift.

Despite these attempts at improving our statistical leverage, we are
unable to find any effect whatsoever. We strongly exclude anything of
the magnitude seen by GSW05. Our results therefore appear to rule out
this idea as a potential model for the effects seen in the
simulations.

This should not take us entirely by surprise, since the results of the
previous section lead us to expect that any effect should be larger for the
high-mass than for low-mass haloes. This is opposite to the numerical
result. We conclude that although a formal dependence should be
present, it is far too small to measure and cannot explain the results
of GSW05 and CGW06.

\subsection{Clustering and progenitor pancakes and filaments}

Since our multi-dimensional extension of the excursion set approach is
unable to reproduce the required correlation between clustering and
formation history, we now turn to the other possibility laid out in
section \ref{correlations}, namely that the correlation may reflect a
dependence of formation history on the properties of progenitor
pancakes and filaments.  In particular we study the correlation
between the masses of these progenitor structures and the large-scale
environment of the $z=0$ halo.
\begin{figure}
\centerline{\epsfxsize=9.0cm\epsffile{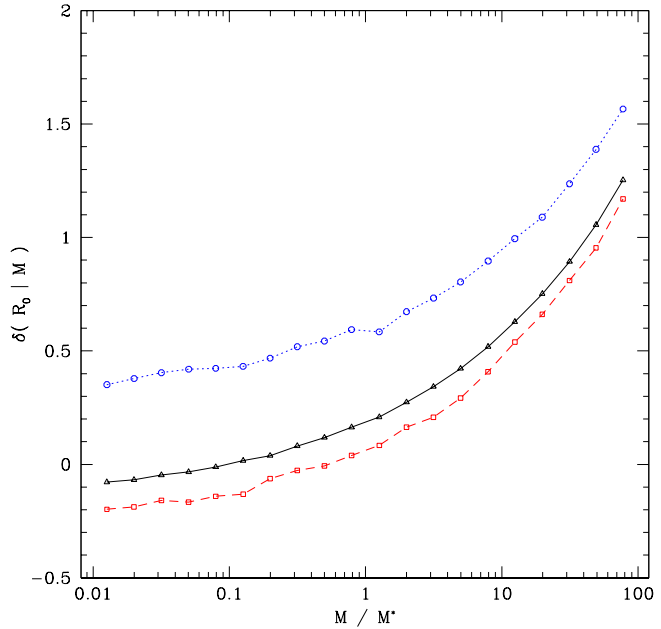}}
\caption{ Average density at large scale ($R_0 \sim 10 h^{-1}$Mpc)
around a $z=0$ halo, as a function of halo mass (black curve) together
with the same quantity for halo subsets defined so that the progenitor
pancakes at $z=2$ lie in the upper (blue curve) and lower (red curve)
10\% tails of the pancake mass distribution.  }
\label{dEnvMass}
\end{figure} 

\begin{figure}
\centerline{\epsfxsize=9.0cm\epsffile{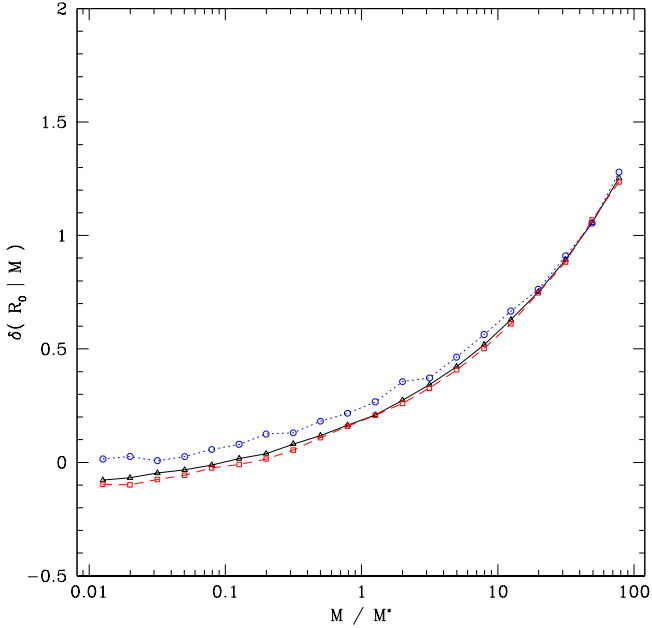}}
\caption{ Average density at large scale ($R_0 \sim 10 h^{-1}$Mpc)
around a $z=0$ halo, as a function of halo mass (black curve) together
with the same quantity for halo subsets defined so that the progenitor
filaments at $z=2$ lie in the upper (blue curve) and lower (red curve)
10\% tails of the filament mass distribution.  }
\label{dEnvMass_filaments}
\end{figure} 

Figure \ref{dEnvMass} plots average large-scale ($\sim 10 h^{-1}$Mpc)
environment overdensity as a function of halo mass, and compares this
with the values found using only those haloes whose progenitor
pancakes at $z=2$ lie in the upper or in the lower $10\%$ tail of the
pancake mass distribution.  There is a strong correlation between this
large-scale overdensity and the mass of the progenitor pancake, with
haloes in denser environments tending to have more massive progenitor
pancakes. The strength of this effect is greatest for low-mass haloes.

Fig. \ref{dEnvMass_filaments} shows a similar plot of large-scale
environment overdensity as a function of halo mass, but now sorting
haloes by progenitor filament mass at $z = 2$. A dependence is again
clear, but is much weaker than was the case for progenitor pancakes.
The smallness of the effect in this case is due to the fact that very
few haloes were embedded in higher mass filaments at $z=2$. Only for
these haloes is any correlation between environment and progenitor
filament mass to be expected, as explained in
sec. (\ref{correlations}).

We have established a clear correlation between environment
overdensity at large scales and the mass of progenitor pancakes and
filaments. 
If
the formation history of a halo is somehow affected by the size of its
progenitor pancakes and filaments, this correlation will induce a
correlation between clustering and formation history. Intriguingly the
effect is largest for low-mass haloes and becomes quite weak for the
most massive haloes. This is precisely the behaviour found by GSW05
and CGW06, where the measured effects are only strong for haloes less
massive than $\sim 10^{13}M\odot$. In the schematic treatment of this
section, the mass at which the dependence becomes small is dependent
on whether we consider filaments or pancakes, and on the redshift at
which we consider them to be relevant.

It is beyond the scope of the current discussion to present a detailed
model for the dependence of halo formation history on the properties
of progenitor pancakes and filaments. Such a connection is intuitively
appealing, however, and we expect this to be an area where further
analytic modelling will be very fruitful.


\subsection{Pre-heating by pre-virialization?}
Our barriers for collapse along one or two axes also allow us to
address a recent claim by \citet{Mo05} that the material which now
resides in low-mass haloes ($M \la 10^{12} M_\odot$) was pre-heated by
the collapse of larger-scale pancakes (typical mass $M \sim 5\times
10^{12}M_\odot h^{-1}$) at $z\sim 2$, and that as a result the gas
component of many of these haloes failed to cool and form stars. The
analytic argument underlying this claim was based on a simplified
calculation which specifically failed to include the conditional
distribution of the masses of progenitor pancakes for haloes of given
$z=0$ mass. We are now well equipped to carry out a more precise
calculation of the relevant quantities.

\begin{figure}
\centerline{\epsfxsize=9.0cm\epsffile{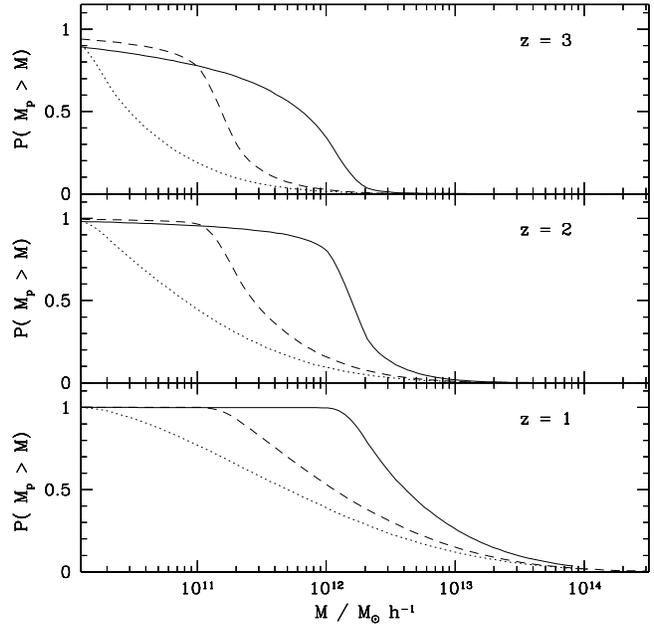}}
\caption{ Cumulative distribution of progenitor pancake masses at
several redshifts for haloes of $z=0$ mass $10^{12} M_{\odot}$
(solid), $10^{11} M_\odot$ (dashed) and $10^{10} M_{\odot}$
(dotted). At $z=2$ only $\sim 10\%$ of $10^{12}M_\odot$ haloes had
progenitor pancakes of the size suggested by \citet{Mo05}. For
$10^{11} M_\odot$ and $10^{10} M_\odot$ haloes this fraction is
significantly smaller. Note that almost all haloes in these mass ranges were
embedded in larger mass pancakes at $z=1$. At $z=3$ this fraction is
$30\%$ for the $10^{12}M_\odot$ haloes but is still large ($\sim80\%$)
for the smaller mass ranges.  }
\label{massfunctions_sheets}
\end{figure}

\begin{figure}
\centerline{\epsfxsize=9.0cm\epsffile{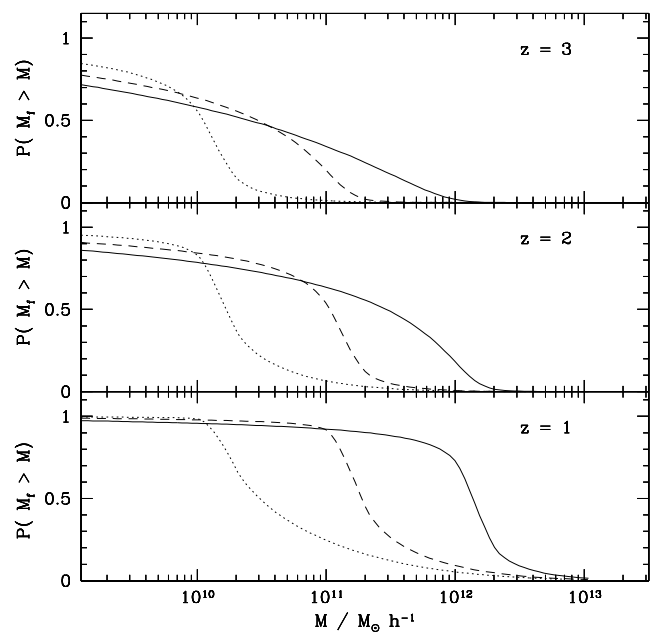}}
\caption{ Cumulative distribution of progenitor filament masses at
several redshifts for haloes of $z=0$ mass $10^{12} M_{\odot}$
(solid), $10^{11} M_{\odot}$ (dashed) and $10^{10} M_\odot$ (dotted).
}
\label{massfunctions_filaments}
\end{figure} 

To examine this claim we pick out haloes in narrow mass ranges around
$M = 10^{12} M_\odot$, $M = 10^{11} M_\odot$ and $M = 10^{10}
M_\odot$, and we look at the mass distribution of their progenitor
pancakes and filaments at redshifts 1, 2, and 3. These cumulative
distributions are shown in figs. (\ref{massfunctions_sheets})
(pancakes) and (\ref{massfunctions_filaments}) (filaments).  They show
that many low-mass haloes were indeed embedded in larger mass
structures at redshifts $z \sim 1 - 2 $, in qualitative agreement with
\citet{Mo05}.

Quantitatively however, the mechanism looks less promising. Central to
the argument of \citet{Mo05} was the assumption that most low-mass
haloes were embedded in pancakes of mass $M \sim 5 \times 10^{12}
M_\odot$ at $z\sim 2$. This was required to ensure sufficient
preheating of gas to affect later condensation and star
formation. From the middle graph in fig. (\ref{massfunctions_sheets})
it is clear that typical $10^{11} M_\odot$ haloes had progenitor
pancakes of mass $M \sim 2 \times 10^{11} M_\odot $ at $z=2$, and only
a few percent have progenitor pancakes as massive as assumed by
\citet{Mo05}. The situation is even worse for typical $10^{10}
M_\odot$ haloes which were embedded in pancakes of mass $M \sim
7\times 10^{10}M_\odot$ at $z = 2$, nearly two orders of magnitude
less than the required value.

Our results contradict those of \citet{Mo05} in two ways. According to
our analysis, progenitor pancake mass is not independent of halo mass
in the way they suggest, and even for relatively massive ``dwarf
galaxy'' haloes, the progenitor pancakes are typically too small to
generate the temperatures required for suppression of later cooling.
Taking both these discrepancies into account, it seems unlikely that
the \citet{Mo05} mechanism can account for the shallow faint end slope
of the galaxy luminosity function and the relatively small number of
luminous satellites around galaxies like the Milky Way.

\section{Conclusions}
We have used the excursion set formalism to discuss the correlation
between environment and formation history found for haloes in
large-scale simulations of cosmic structure formation.  We implemented
a multi-dimensional generalisation of the excursion set approach which
assumes halo formation to follow the triaxial collapse model of
\citet{BM96} and which does not require further simplifying assumptions of
the kind introduced by \citet{SMT01} to reduce the system to a
1-dimensional barrier crossing problem.  A correlation between halo
formation history and halo environment is expected in this model
\citep{Wang06} but we have shown here that it is very weak.  We
\emph{did}, however, demonstrate that there is a measurable
correlation between environment density and the pre-collapse shape of
a halo.


It appears that haloes in dense regions have slightly more extreme
axial ratios than similar mass objects in underdense environments.
This could potentially be studied with N-body simulations, although we
caution that our definition of ellipticity, eqn.(\ref{ellipticity}),
while mathematically convenient, is not simply related to the
measurable shape of nonlinear, quasi-equilibrium haloes. To address
this issue properly, we suggest using large N-body simulations such as
the Millennium Simulation \citep{Springel05}. It would be feasible to
follow particles which are part of $z=0$ haloes back to the initial
density field. With a large enough sample of haloes one could then
determine how the final shapes of halos are related to the properties
of their initial, linear ``patches''.

With straightforward definitions of environment density and
halo formation time our multi-dimensional random walks are
inconsistent with the environment-formation history correlations found
by GSW05 and CGW06. We have, however, discovered a different mechanism which can
induce a dependence of some aspects of halo formation on halo
environment.  Crucial to our analysis of this mechanism are the
collapse barriers for haloes, filaments and sheets introduced in
section (\ref{barriers}).  We showed that many low-mass $z=0$ haloes
were embedded in larger-mass progenitor pancakes at redshifts $z \sim
1-2$, and that the mass of these progenitors correlates with the
density of the environment surrounding the $z=0$ haloes. Thus, if the
formation history of $z=0$ haloes is affected by the mass of their
high-redshift progenitor pancakes, we can expect an induced dependence
of formation history on environment.  Although we do not attempt to
model this, we show that sorting haloes by the mass of their
progenitor pancakes at $z \sim 1$ to 3 gives a variation of mean
environment with pancake mass which is of similar size and has a
similar dependence on halo mass to the variation found in simulations
when haloes are ranked by their formation time.  Future efforts should
result in a more detailed model for the connection between formation
history and the masses of progenitor pancakes and filaments.

A very recent paper by \citet{Wang06} suggests that the age dependence
of clustering observed by GSW05 is really caused by mass stripping due
to the large scale tidal field, which makes small old haloes appear
less massive than their initial density field would suggest. This
might provide an intuitive explanation for the effect, and is an
example of an effect which could be connected to our scenario.

By studying conditional mass functions for the sheets and filaments in
which the material of low-mass $z=0$ haloes was embedded at early
times, we found typical masses for progenitor pancakes at $z=2$ which
are significantly smaller than suggested by \citet{Mo05}.  In
addition, these progenitor masses depend strongly on the mass of the
final halo. These results suggest that preheating due to the collapse
of progenitor pancakes is insufficient to suppress later cooling and
condensation of gas.  It is
nevertheless interesting that the sheet mass function is more strongly
peaked than its halo counterpart, so that the majority of low-mass
haloes were indeed embedded in more massive pancakes at $z \sim 2$.

In summary we argue that this work may provide an interesting and
intuitive route to understanding how environment influences halo
formation. It is important that analytic methods are refined to
account at least qualitatively for the relatively strong effects seen
in N-body simulations, and our results provide a first step in this
direction. To build on the approach presented here it will necessary
to model in more detail how the structural elements of the cosmic web
influence the formation of haloes within them. We hope our results
will motivate such efforts.

\appendix

\section[]{The Ellipsoidal Collapse Model}
For convenience we here review the most important features of the
ellipsoidal collapse model of BM96.

Starting from a homogeneous ellipsoidal patch with density contrast
$\delta$, the general equation that governs the collapse of an
initially spherical surface along principal axis $k$ is
\be
{d^2A_k \over dt^2} = -4\pi G \bar{\rho} A_k \left[ {1+\delta \over 3}
  + { b_k' \over 2}\delta +\lambda_k' \right]
\ee
where $b_k' \delta / 2$ and
$\lambda_{k}'$ denote the interior and exterior tidal forces
respectively. The interior tidal forces can be calculated for a
constant density patch from the potential theory of homogeneous
ellipsoids (cf. Binney \& Tremaine 1987), and are given by:
\be
b_k'=-\frac{2}{3}+\prod_{i=1}^{3}a_i\int_0^{\infty}\,\frac{d\tau}{(a_k^2+\tau)\prod_{j=1}^3\sqrt{a_j^2+\tau}}.
\ee
The linear approximation for external tides is:
\be
\lambda'_{ext k} = {a \over a_i}\left[\lambda_k(t_i)-{\delta_i \over 3}
\right]
\ee
where the ${\lambda_k}$s are the initial eigenvalues of the
deformation tensor.
The evolving quantity $A_k$ relates to the boundary radius of the
Lagrangian sphere, $R$ through $A_k(t)=a_k(t)R$, so that the $a_k$ can
be thought of as the anisotropic generalisation of the standard cosmic
expansion factor.
We should note here that these equations are valid only in
cosmological models with vanishing cosmological constant -- a
discussion of the modifications for a non-zero cosmological constant
can be found in BM96.

The initial conditions are set by the Zel'dovich
approximation
\bea
A_k(t_i) &=& a_i \left[1 - \lambda_k(t_i) \right] \\
\dot{A}_k(t_i) &=& H(t_i)A_k(t_i) - a_iH(t_i) \lambda_k(t_i)
\eea

The differential equations A1 are evolved for all axes $k$ until a
value of $A_k = 0.177 a$ has been reached, after which the individual
$A_k$ are held constant. The factor of $0.177$
results from the requirement that the virial density contrast of 179
obtained from spherical top hat calculations is reproduced.

\section[]{Invariants and Variables} \label{invariants}
The three rotational invariants of the $3\times 3$ deformation tensor are
\bea
I_1 &=& \lI + \lII + \lIII \\
I_2 &=& \lI \lII + \lII \lIII+ \lIII \lI \\
I_3 &=& \lI \lII \lIII.
\eea
They can all be easily expressed in terms of the ${ y_i }$s. It would
therefore be useful if we could express our barrier variables
in terms of these invariants.

The first barrier variable is naturally the density contrast, $\delta =
I_1 = y_1$. The second variable chosen by CL01, is a
combination of the first two invariants 
\be
r^2 = 3/2 I_1^2 - 2 I_2 = {1 \over 3} \sum_{i\ne j} (\lambda_i-\lambda_j)^2
\ee
which satisfies the condition that it is \emph{independent} of
$y_1$. This is important both since we need to fit the density
contrast at collapse to the barrier parameter, and because it means that it
vanishes in the spherical case. (The spherical case is equivalent to
$y_{i \ne 1}=0$.)

For the third barrier variable a similarly convenient combination of the
three invariants would be
\be
u^3 = {1\over 9} \left(2 I_1^3 - 9I_1I_2 + 27I_3 \right) 
\ee 
which is also independent of $y_1$ and thus also vanishes in the spherical 
case.

The two variables $r^2$ and $u^3$ may be obtained directly from the
$y$s as follows: 
\bea 
r^2 &=& {2\over 15}( y_2^2+y_3^2+y_4^2+y_5^2+y_6^2 )\\
u^3 &=& {1\over
  9}{1\over 25}(-3\sqrt{5}\sqrt{3}y_5^2y_2-3\sqrt{5}y_5^2y_3+6\sqrt{5}y_6^2y_3 \nonumber \\
& & +3\sqrt{5}\sqrt{3}y_4^2y_2-3\sqrt{5}y_4^2y_3+6
\sqrt{5}y_2^2y_3 \nonumber \\
& &+2 -\sqrt{5}y_3^3-6 \sqrt{5}\sqrt{3}y_4y_5y_6) 
\eea

where we have used the explicit formulae in \citet{LCL02}(LCL02). We thus have two
variables which are expressed in terms of the rotational invariants
and which are independent of the density contrast. 
The question is whether a barrier in these two variables is capable of
capturing the behaviour of the ellipsoidal collapse model, and here we
stumble upon a problem:

Although the transformation involved is one-to-one it does not
guarantee that it will allow one variable to be a single valued
function of the two others. Indeed it is easy to show that
transforming from eigenvalue space to ${\delta, r^2, u^3}$
occasionally yields identical $r$ and $u$ values for different
$\delta_c$s - thereby preventing an expression of $\delta_c$ as a
single valued function of the other two variables.

It proves significantly more accurate to express the barrier variables as
suitable orthogonal linear combinations of the eigenvalues as follows 
\bea
\delta &=& \lI + \lII + \lIII \\
v & = & -\lI + \lII\\
w & = & -\lI -\lII +2 \lIII .
\eea
The variables $v$ and $w$ both vanish in the spherical case, are
independent of $\delta$ and of each other, and are always positive for
ordered eigenvalues $\lI < \lII < \lIII$. Since we prefer to work with a
barrier which reproduces the ellipsoidal collapse model of BM96
as well as possible, we use this set of variables throughout our work.   

The joint probability distribution can be found by manipulation of
eqn. (\ref{doroshkevich})
\bea
p(\delta, v, w|\sigma) &=& {1 \over 3 }{15^3 \over 2^6}{1 \over
  \sqrt{5}\pi\sigma^6} \exp\left( - {1 \over 2\sigma^2} (\delta^2+ {15\over 4}
v^2 + {5 \over 4} w^2) \right) \nonumber \\
& & \times (w^2-v^2) v,
\label{doroshkevich_new}
\eea
where the limits are 
\bea
-\infty & <  \delta  < & \infty \\
0 &\le  v  \le & w  \\
0 & \le w  < & \infty
\eea
The expectation values of $v$ and $w$ for a given $\sigma$ are
\bea
\langle v \rangle & = & {3 \sigma \over \sqrt{10 \pi}} \\
\langle w \rangle & = & {9 \sigma \over \sqrt{10 \pi}}.
\eea
These are the values used to obtained the average barrier
functions in figure \ref{randomWalk} in the main text.

\bsp

\label{lastpage}

\end{document}